\documentclass[aps,prl,twocolumn,superscriptaddress,floatfix,amsmath,amssymb,nofootinbib,longbibliography]{revtex4-2}

\usepackage{graphicx}
\usepackage{xcolor}
\usepackage{dcolumn}
\usepackage{bm}
\usepackage[T1]{fontenc}
\usepackage{orcidlink}
\usepackage{hyperref}
\hypersetup{
colorlinks = true,
citecolor  = blue,
linkcolor  = blue,
urlcolor  = blue
}
\begin{document}

\title{\textit{Ab Initio} Calculations of Beta-Decay Half-Lives for $N=50$ Neutron-Rich Nuclei}

\author{Zhen Li\,\orcidlink{0000-0003-2786-7272}}
\email{zhen.li1@tu-darmstadt.de}
\affiliation{Technische Universit\"at Darmstadt, Department of Physics, D-64289 Darmstadt, Germany} 
\affiliation{ExtreMe Matter Institute EMMI, GSI Helmholtzzentrum f\"ur Schwerionenforschung GmbH, D-64291 Darmstadt, Germany}
\affiliation{Max-Planck-Institut f\"ur Kernphysik, Saupfercheckweg 1, D-69117 Heidelberg, Germany}

\author{Takayuki Miyagi\,\orcidlink{0000-0002-6529-4164}}
\email{miyagi@nucl.ph.tsukuba.ac.jp}
\affiliation{Center for Computational Sciences, University of Tsukuba, 1-1-1 Tennodai, Tsukuba 305-8577, Japan}

\author{Achim Schwenk\,\orcidlink{0000-0001-8027-4076}}
\email{schwenk@physik.tu-darmstadt.de}
\affiliation{Technische Universit\"at Darmstadt, Department of Physics, D-64289 Darmstadt, Germany} 
\affiliation{ExtreMe Matter Institute EMMI, GSI Helmholtzzentrum f\"ur Schwerionenforschung GmbH, D-64291 Darmstadt, Germany}
\affiliation{Max-Planck-Institut f\"ur Kernphysik, Saupfercheckweg 1, D-69117 Heidelberg, Germany}

\begin{abstract}
Beta-decay rates of extreme neutron-rich nuclei remain largely unknown experimentally, while they are critical inputs for $r$-process nucleosynthesis. We present first \textit{ab initio} calculations of total beta-decay half-lives, with a focus on $N=50$ nuclei. Starting from nuclear forces and currents based on chiral effective field theory, we use the in-medium similarity renormalization group to consistently derive valence-space Hamiltonians and weak operators, from which we calculate the nuclear states involved and the Gamow-Teller transition strengths, without phenomenological adjustments. In addition, we explore effects of first-forbidden contributions. Our results show that the inclusion of two-body currents increases the total half-lives, which then show good agreement with the existing experimental data, thereby validating the predictive capability of our approach.
\end{abstract}

\maketitle

\textit{Introduction}---The rapid neutron-capture process (or $r$-process) is responsible for generating over half of the heavy elements beyond iron in our Universe~\cite{Kajino2019,Cowan2021,Arcones2023}. Although essential for $r$-process simulations, the beta-decay half-lives of neutron-rich nuclei along magic neutron numbers $N=50,\ 82,\ 126,\ldots$ (known as $r$-process waiting point nuclei) remain largely unknown experimentally (see, e.g., Ref.~\cite{Mumpower2016}). For these extreme neutron-rich nuclei, $r$-process calculations depend on theoretically predicted half-lives. Evaluating the half-lives of $r$-process waiting point nuclei is, therefore, critically needed.

Existing half-life calculations for $r$-process waiting point nuclei are largely based on the quasiparticle random-phase approximation (QRPA)~\cite{Engel1999,Borzov2003,Moller2003,Marketin2007,Mustonen2016,Marketin2016,Ney2020,Robin2024} or the nuclear shell model~\cite{MartinezPinedo1999,Cuenca-Garca2007,Suzuki2012,Zhi2013,Yoshida2018,Kumar2024} using phenomenological interactions and corrections. For example, it is known that QRPA underestimates many-body correlations~\cite{Marketin2016}, while a recent extension of QRPA to account for the coupling between single-particle and collective degrees of freedom yielded improved results~\cite{Robin2024}. On the other hand, the nuclear shell model provides a better agreement with the available data~\cite{Cowan2021}. However, phenomenological shell-model calculations require \textit{ad hoc} adjustments by quenching the axial-vector coupling $g_A$ to reproduce experimental Gamow-Teller (GT) strengths. 

In the past decades, the development of nuclear forces from chiral effective field theory (EFT)~\cite{Epelbaum2009,Machleidt2011} combined with powerful many-body approaches~\cite{Hergert2020_review,Hebeler2021} has made it possible to perform systematically improvable \textit{ab initio} calculations. Recently, the $g_A$ quenching puzzle in GT transitions was successfully explained in \textit{ab initio} calculations~\cite{Gysbers2019} by taking into account many-body correlations and consistent two-body (2B) currents from chiral EFT.

In this Letter, we present first \textit{ab initio} calculations of total beta-decay half-lives using the valence-space in-medium similarity renormalization group (VS-IMSRG)~\cite{Tsukiyama2011,Tsukiyama2012,Hergert2016_IMSRG,Stroberg2017,Stroberg2019_review_Heff}. We focus on the astrophysically relevant $N=50$ waiting point nuclei, which are also an active target of experiments~\cite{Hosmer2005,Hosmer2010,Xu2014,Tolosa2025}. Our results are based on chiral nucleon-nucleon (NN) and three-nucleon (3N) interactions, with a particular focus on the role of 2B currents in the many GT transitions for the total rates. We find good agreement with experiment without phenomenological adjustments and make predictions for the half-lives at $Z=24-26$, exploring also the effects of first-forbidden (FF) contributions.

\textit{Theoretical framework}---The total beta-decay half-life $T_{1/2}$ of a nucleus in the initial state i (the ground state in our case) is obtained by summing over the partial half-lives to all possible final states f: $T_{1/2}^{-1} = \sum_\text{f} t_\text{fi}^{-1}$. For allowed GT transitions, $t_\text{fi}$ reads~\cite{Suhonen07} (using $\hbar=c=1$) as follows:
\begin{equation}
t_\text{fi}^{-1} = \frac{B({\rm GT})}{\kappa} \int_1^{W_0} dW \, F(Z,W) \, p_e \, W (W_0-W)^2 \,,
\label{eq:integral}
\end{equation}
with electron energy $W$, maximum electron energy $W_0$, and electron momentum $p_e=\sqrt{W^2-1}$, all in units of electron mass, and $\kappa = 6144.48 \pm 3.7$~s~\cite{Hardy2020}. $F(Z,W)$ is the Fermi function, which takes into account the Coulomb distortion of the electron wave function near the final nucleus (with proton number $Z$) as well as the finite nuclear size~\cite{Behrens1971,Behrens1982}. 
We neglect the contributions from Fermi transitions, as they are expected to be small because they predominately involve isobaric analog states, which are very high-lying for the initial states considered in this Letter. Therefore, we only consider allowed GT transitions and will discuss FF transitions later. The GT transition strength $B({\rm GT})$ is given by
\begin{equation}
B({\rm GT}) = \frac{1}{(2J_\text{i}+1)} \bigl| \langle\text{f}||{\rm GT}||\text{i}\rangle \bigr|^2 \,,
\end{equation}
where $J_\text{i}$ is the total angular momentum of the initial state. The GT operator is given by the spatial axial-vector current ${\bf J}$. Since the momentum transfer ${\bf q}$ is very low in beta decays, we can evaluate the axial-vector current at vanishing momentum transfer. For $\beta^{-}$ decay, ${\rm GT} = {\bf J}_{x} + i {\bf J}_{y}$ with isospin components ${\bf J}_{x}$ and ${\bf J}_{y}$. Up to next-to-next-to-leading order (N$^2$LO) and at $|{\bf q}|=0$ the axial-vector current has contributions from the leading one-body (1B) and leading 2B currents~\cite{Park2003,Krebs2017,Klos2017,Hoferichter:2020osn},
${\bf J} = {\bf J}_{\rm 1B} + {\bf J}^{\rm ct}_{\rm 2B} + {\bf J}^{\rm 1\pi}_{\rm 2B}$, with
\begin{align}
{\bf J}_{{\rm 1B},i} &= \frac{g_{A}}{2} \, \bm{\sigma}_i \bm{\tau}_i \,, \\
{\bf J}^{\rm ct}_{\rm 2B,12} &= -d_1(\bm{\sigma}_1\bm{\tau}_{1} + \bm{\sigma}_2\bm{\tau}_{2}) -d_2 \, \bm{\sigma}_\times \bm{\tau}_\times \,, \\
{\bf J}^{\rm 1\pi}_{\rm 2B,12} &= -\frac{g_A}{2F_\pi^2} \frac{\bm{\sigma}_2 \cdot {\bf k}_2}{{\bf k}_2^2+ m_\pi^2} \biggl[ 2 c_3 \, \bm{\tau}_2 \, {\bf k}_2 \nonumber \\
&\quad+ \left(c_4+\frac{1}{4m}\right) \bm{\tau}_\times (\bm{\sigma}_1 \times {\bf k}_2) \biggr] + (1 \leftrightarrow 2) \,.
\end{align}
The 1B current depends on the axial coupling $g_A=1.27$ and the spin and isospin operators of the $i$th nucleon, $\bm{\sigma}_i$ and $\bm{\tau}_i$.
The short-range contact (ct) 2B current depends on $d_1, d_2$, which upon antisymmetrization is given by the single contribution $c_D = -4(d_1+2d_2)F_\pi^2 \Lambda_{\chi}$~\cite{Park2003,Gazit2009,Klos2017,Baroni2018} taken consistently from the N$^2$LO 3N interaction. Here, $F_\pi = 92.2$\,MeV is the pion decay constant and $\Lambda_{\chi} = 700$\,MeV. Moreover, $\bm{\sigma}_\times = \bm{\sigma}_1 \times \bm{\sigma}_2$ and similarly for the isospin operator $\bm{\tau}_\times$. The one-pion-exchange (${\rm 1\pi}$) 2B current depends on the couplings $c_3, c_4$ that we take consistent with the NN and 3N interactions. ${\bf k}_i = {\bf p}_i'-{\bf p}_i$ is the difference of the final and initial $i$th nucleon momenta, and $m_\pi=138.04$\,MeV and $m=938.92$\,MeV are the (averaged) pion and nucleon mass, respectively. The 2B currents are regularized with nonlocal regulators of the same form as used for the 3N interactions.

For calculating the structure of the initial and final nuclear states and the GT transitions, we use the VS-IMSRG. The IMSRG~\cite{Tsukiyama2011,Hergert2016_IMSRG} starts from the full Hamiltonian with NN and 3N interactions, normal ordered with respect to a reference state, and uses flow equations (or equivalently a continuous series of unitary transformations) to decouple particle-hole excitations from the reference state. In the case of the VS-IMSRG~\cite{Tsukiyama2012,Stroberg2017,Stroberg2019_review_Heff} the flow equations are used to decouple a valence-space Hamiltonian from excitations beyond the valence space. After the VS-IMSRG evolution, the valence-space Hamiltonian is subsequently diagonalized to include the low-lying correlations in the valence space.

We use the 1.8/2.0~(EM) NN+3N interaction~\cite{Hebeler2011_EM1.8_2.0} for our main result, as it has been successfully used for energies (see, e.g., Ref.~\cite{Stroberg2021}) and for calculations of beta-decays~\cite{Gysbers2019}. The 1.8/2.0~(EM) interaction is fit to NN scattering, the $^3$H energy, and the $^4$He radius. To test the sensitivity to the input Hamiltonian, we also consider the $\Delta$N$^2$LO$_{\rm GO}$~(394) NN+3N interaction~\cite{Jiang2020}, which is fit to lower energy NN scattering, $A=3,4$ properties, and informed by heavier nuclei and nuclear matter. For each nucleus, we use a Hartree-Fock basis calculated from the NN+3N interaction in 15 major harmonic-oscillator shells [$e_{\rm max}={\rm max}(2n+l)=14$] with $\hbar\omega=16$\,MeV and three-body matrix elements restricted to $E_{\rm 3max}=24$. This is sufficient for converged results for $A<100$~\cite{Miyagi2022_E3max}. The NN and 3N interaction and the GT 2B current matrix elements are obtained using the \texttt{NuHamil} code~\cite{Miyagi2023}.

In this Letter, we explore the $N=50$ waiting point nuclei from $A=74-82$. We take the valence space to consist of the $\{0f_{7/2}, 0f_{5/2}, 1p_{3/2}, 1p_{1/2}\}$ proton orbitals and the $\{0f_{5/2}, 1p_{3/2}, 1p_{1/2}, 0g_{9/2}, 1d_{5/2}\}$ neutron orbitals on top of a $^{48}$Ca core. Compared to~\cite{Zhi2013}, we include the $1d_{5/2}$ neutron orbital, given that this is the lowest above $0g_{9/2}$ in our VS-IMSRG calculations, and because neutron particle-hole excitations have important contributions to the first $2^+$ state of $^{78}$Ni~\cite{Hagen2016}.

\begin{figure}[t!]
\centering
\includegraphics[width=\columnwidth]{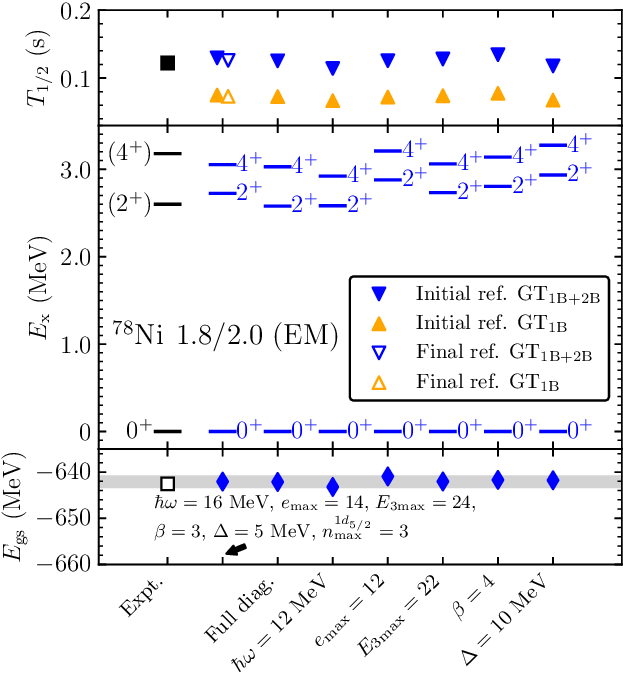}
\caption{Total beta-decay half-life (top), low-lying excited states (middle), and ground-state energy (bottom panel) of $^{78}$Ni calculated using the VS-IMSRG with the 1.8/2.0~(EM) NN+3N interaction. We compare our main result (second column) to the experiment (left column)~\cite{Xu2014,nndc}. Note that the experimental ground-state energy is not known and taken from the adopted AME 2020 value~\cite{AME2020}. The third to eighth columns show the very weak dependence of our main result to relaxing the occupation constraint for the $1d_{5/2}$ orbital, to $\hbar \omega$ variation, to the single-particle basis $e_{\rm max}$, to the $E_{\rm 3max}$ restriction of 3N matrix elements, and to the choice of $\beta$ and $\Delta$, respectively. For $T_{1/2}$ we show the impact of 2B currents (from orange up to blue down triangles) and the negligible dependence on evolving the current operators with respect to the initial or final reference nucleus.}
\label{fig:Ni78}
\end{figure}

For the VS-IMSRG, we take the state-of-the-art VS-IMSRG(2) truncation with ensemble normal ordering~\cite{Stroberg2017}. The VS-IMSRG(2) keeps up to normal-ordered 2B contributions to the interactions and currents in the evolution. Three-body-operator contributions are expected to contribute $1\%$-$2\%$ to the correlation energy~\cite{Heinz2025} [for the $^{78}$Ni ground-state energy this corresponds to 2-4 and 3-6\,MeV for the 1.8/2.0~(EM) and $\Delta$N$^2$LO$_{\rm GO}$~(394) interactions, respectively]. We use the arctan variant of the White generator with $\Delta=5$\,MeV for our multishell valence space~\cite{Miyagi2020}, and the VS-IMSRG unitary transformation is realized via the Magnus formulation~\cite{Morris2016}. Moreover, we use a modified Hamiltonian $H' = H+ \beta H_{\rm c.m.}$ with center-of-mass (c.m.) Hamiltonian $H_{\rm c.m.}$ and $\beta=3$ to remove spurious c.m. effects~\cite{Lawson,Miyagi2020}. The 1B and 2B current operators are consistently transformed, keeping up to normal-ordered 2B contributions. Finally, the total beta-decay half-lives are computed using the Lanczos strength-function method~\cite{Haxton2005,CaurierRMP} to efficiently generate the final states. The VS-IMSRG calculation is performed using the \texttt{IMSRG++}~\cite{imsrgpp} and \texttt{KSHELL} codes~\cite{Shimizu2019}. 

\textit{Results for $^{78}$Ni}---We first consider $^{78}$Ni to test the structure calculation and quantify the uncertainties from our theoretical choices. Figure~\ref{fig:Ni78} shows our VS-IMSRG results for the total beta-decay half-life, the $0^{+}$ ground-state energy, and the $2^{+}$ and $4^{+}$ excitation energies based on the 1.8/2.0~(EM) interaction. Our main results (also for the other $N=50$ nuclei) are with $\hbar\omega=16$\,MeV, $e_{\rm max}=14$, $E_{\rm 3max}=24$, $\beta=3$, $\Delta=5$\,MeV. In addition, we use a truncation to limit the number of neutrons in the $1d_{5/2}$ orbital to $n_{\rm max}^{1d_{5/2}}=3$, given that the valence-space dimensions for most nuclei studied in this work are larger than $10^9$. In Fig.~\ref{fig:Ni78}, we show in detail that the dependence on these choices is very small, including relaxing the $n_{\rm max}^{1d_{5/2}}$ truncation to perform the full diagonalization, which is possible for $^{78}$Ni. We use this variation to estimate an uncertainty for the ground-state energy of 2.3\,MeV (shown as gray band in Fig.~\ref{fig:Ni78}) with a similar estimate of 2.2\,MeV for $^{78}$Cu. However, the uncertainty estimated from this simple variation is likely underestimated, and the full contributions from Hamiltonian uncertainty and many-body uncertainty are not taken into account. We note that the spectrum is similarly well-reproduced as for other VS-IMSRG calculations~\cite{Taniuchi2019,Tichai2023DMRG}, where a different valence space (not applicable for beta decays) was used. 

Moreover, Fig.~\ref{fig:Ni78} clearly shows the important impact of 2B currents, which is to decrease the GT strength and thus increase the half-life in all cases. This 2B current effect is significant for all theoretical choices in Fig.~\ref{fig:Ni78} and does not depend on whether the initial reference nucleus is used for the operator evolution (for our main results) or the final nucleus. Once 2B currents are included, the total half-life agrees very nicely with experiment, without the need for phenomenological adjustments. For a complete uncertainty quantification, we need to take into account the uncertainty due to the input Hamiltonian and operators. This will be explored in the following, albeit in a more limited way, using the $\Delta$N$^2$LO$_{\rm GO}$~(394) interaction and studying the impact of FF contributions.

\begin{figure}[t!]
\centering
\includegraphics[width=\columnwidth]{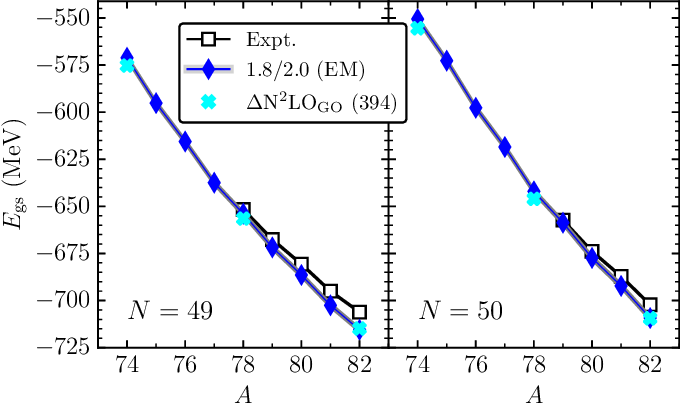}
\caption{Ground-state energies of neutron-rich $N=49$ (left) and $N=50$ isotones (right panel) calculated from the VS-IMSRG based on the 1.8/2.0~(EM) and $\Delta$N$^2$LO$_{\rm GO}$~(394) interactions and in comparison with experiment~\cite{nndc}. The gray band for the 1.8/2.0~(EM) interaction estimates the uncertainty from the model-space convergence and does not include interaction or IMSRG(3) uncertainties.
\label{fig:Egs}}
\end{figure}

\textit{Results for $N$=\,50 waiting point nuclei}---In Fig.~\ref{fig:Egs} we show the ground-state energies of the $N=49$ and $N=50$ isotones for the two Hamiltonians considered. Our calculations mildly overestimate the ground-state energies where experimental data exists, by $1\%$ for the worst case for the 1.8/2.0~(EM) interaction. This is within the VS-IMSRG uncertainties discussed above.

\begin{figure}[t!]
\centering
\includegraphics[width=\columnwidth]{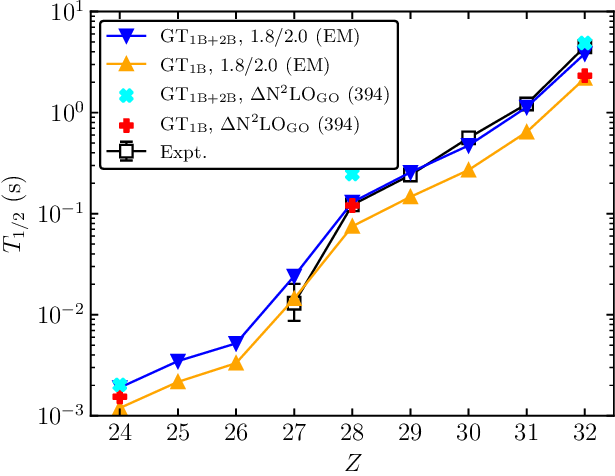}
\caption{Total beta-decay half-lives of $N=50$ waiting point nuclei calculated from the VS-IMSRG, including 1B and 1B+2B current contributions for the 1.8/2.0~(EM) interaction (triangles, our main result) and for the $\Delta$N$^2$LO$_{\rm GO}$~(394) interaction (crosses) in comparison with experiment~\cite{Xu2014,nndc}.
\label{fig:total_half_life}}
\end{figure}

\begin{figure*}[t!]
\centering
\includegraphics[width=0.985\columnwidth]{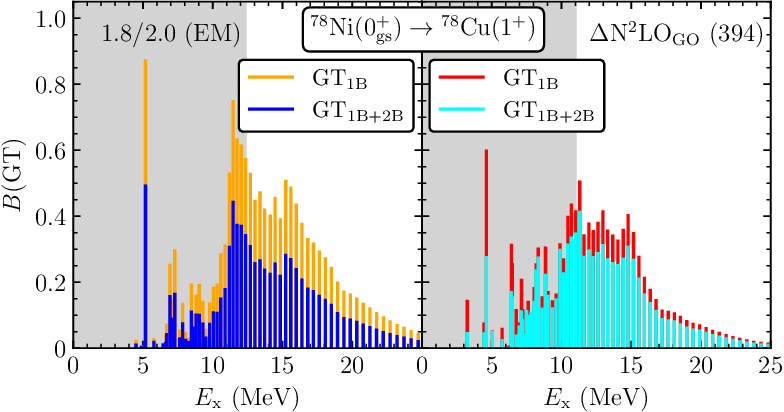}
\hspace*{2mm}
\includegraphics[width=1.02\columnwidth]{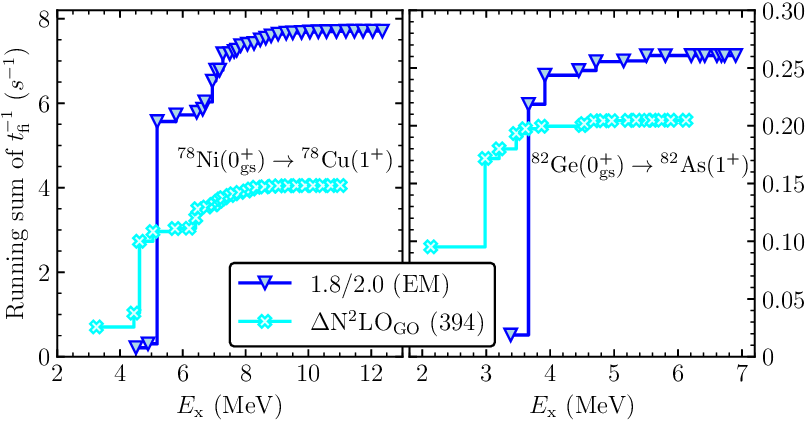}
\caption{Left panel: distribution of GT transition strength $B({\rm GT})$ with and without 2B current contributions for ${^{78}{\rm Ni}(0_{\rm gs}^+)} \rightarrow {^{78}{\rm Cu}(1^+)}$ as a function of excitation energy in the final nucleus. VS-IMSRG results are shown for the 1.8/2.0~(EM) interaction (left, our main result) and for the $\Delta$N$^2$LO$_{\rm GO}$~(394) interaction (right). The shaded area is the kinematically allowed energy region for beta-decay. Right panel: running sums of the inverse partial half-lives for ${^{78}{\rm Ni}(0_{\rm gs}^+)} \rightarrow {^{78}{\rm Cu}(1^+)}$ (left) and ${^{82}{\rm Ge}(0_{\rm gs}^+)} \rightarrow {^{82}{\rm As}(1^+)}$ (right). VS-IMSRG results are shown for both interactions including 2B currents in all cases.
\label{fig:GT_transition_strength}}
\end{figure*}

Our main results for the $N=50$ total beta-decay half-lives calculated from 1.8/2.0~(EM) interaction are presented in Fig.~\ref{fig:total_half_life}. We find that the inclusion of 2B currents leads to longer total half-lives for all $N=50$ nuclei studied, leading to a very good agreement with experiment for $Z=28-32$ and reproducing the trend down to $Z=27$. We emphasize that no adjustments to half-lives or GT transitions have been made. The results are obtained only by using the given Hamiltonian with consistent 2B currents. The effect of the 2B currents can be understood by analyzing the GT transition strength in the left panel of Fig.~\ref{fig:GT_transition_strength}. The inclusion of 2B currents systematically reduces the $B({\rm GT})$ across the whole energy region, resulting in a longer half-life. This is consistent with the general arguments for the reduction of GT contributions from 2B currents~\cite{Menendez2011} and with the studies of the quenching puzzle~\cite{Gysbers2019}. However, so far no {\it ab initio} calculations of total beta-decay half-lives that proceed through many states have been made.

To gain insight into the interaction uncertainty, we have performed calculations with the $\Delta{\rm N}^2{\rm LO}_{\rm GO}~(394)$ interaction for $Z$=\,24, 28, and $32$. The effect of the nucleon excitation to the $\Delta$ isobar in the 2B currents has been taken into account by appropriately shifting the low-energy constants $c_3$ and $c_4$~\cite{Baroni2018}. As shown in Fig.~\ref{fig:total_half_life}, the 2B current contributions also increase the half-lives, with again an overall reduction across the whole energy window in the left panel of Fig.~\ref{fig:GT_transition_strength}. Moreover, we find consistent results for the total half-lives with 2B currents at $Z=24$ and $32$ for both interactions. However, for $^{78}$Ni the $\Delta{\rm N}^2{\rm LO}_{\rm GO}~(394)$ interaction leads to a longer half-life. To analyze this further, we show in the right panel of Fig.~\ref{fig:GT_transition_strength} the running sums of the inverse partial half-lives. For $^{78}$Ni, we observe quite different running sums for two different interactions, as one can expect from the different $B({\rm GT})$ distributions in the left panel. However, the phase space factors given by the integral in Eq.~\eqref{eq:integral} are also different, and the spectrum is more compressed for the $\Delta{\rm N}^2{\rm LO}_{\rm GO}~(394)$ interaction. This shows the intricate interplay of the $B({\rm GT})$ and the phase space factors for the total half-life. For $^{82}$Ge ($Z=32$), the running sum also has a different behavior for the two interactions, but the final total half-life is more similar. This shows that the similarities at $Z=24$ and $32$ may be accidental and assessing the Hamiltonian uncertainty will require significantly more work, necessitating emulators for these complex total half-life calculations.

\begin{figure}[t!]
\centering
\includegraphics[width=\columnwidth]{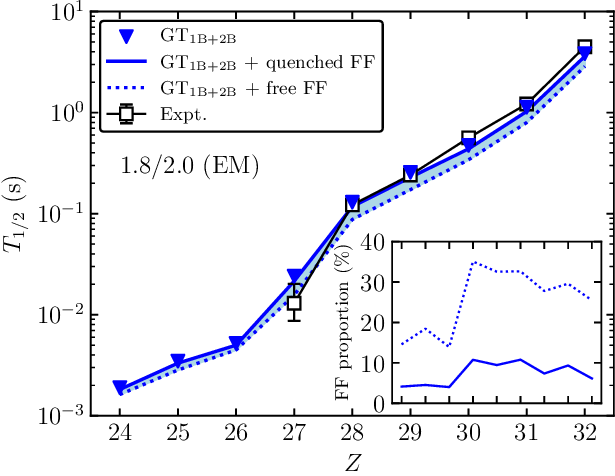}
\caption{Total beta-decay half-lives of $N=50$ waiting point nuclei calculated from the VS-IMSRG for the 1.8/2.0~(EM) interaction with 2B currents compared to experiment~\cite{Xu2014,nndc}. In addition, we show the impact of FF transitions assuming a bare operator (``free FF'') or with phenomenological quenching (``quenched FF''). The inset shows the percentage of FF contributions to the decay rate.
\label{fig:total_half_life_GT_FF}}
\end{figure}

Phenomenological shell-model calculations have shown that FF transitions are non-negligible for the $N=50$ isotones from $Z=24-27$~\cite{Zhi2013}, as the $0f_{7/2}$ proton orbital is not fully occupied in the naive shell-model filling. To explore the role of FF transitions in our calculations, we also investigate FF contributions for our VS-IMSRG calculations. For these, $B({\rm GT})$ in Eq.~\eqref{eq:integral} is replaced by a $W$-dependent shape factor $C(W)$ in the integral~\cite{Behrens1982}. We use the same expressions for $C(W)$ as in~\cite{Zhi2013}. There are nine different operators for the FF contributions. To estimate their effects, we do not evolve them consistently in the VS-IMSRG but include them later, either as bare operator (``free FF'') or with phenomenological quenching from~\cite{Zhi2013}. The total FF transition rate is obtained by summing over the 20 lowest final states for each $J_\text{f}$, which is sufficient for converged results.

Our results for the total beta-decay half-lives with 2B currents and FF contributions are shown in Fig.~\ref{fig:total_half_life_GT_FF}.  We find a reduction due to FF transitions, which is most pronounced for the free FF case. Because the FF operators are not evolved consistently in the VS-IMSRG, we consider the results shown in Fig.~\ref{fig:total_half_life_GT_FF} as an uncertainty range for the half-lives. In our calculations, we find a rather smaller proportion of FF transitions for $Z \leq 26$ and a larger proportion for $Z \geq 27$ due to the additional $1d_{5/2}$ neutron orbital in our valence space compared to~\cite{Zhi2013}.

\textit{Summary and conclusions}---We have presented first \textit{ab initio} VS-IMSRG calculations for the total beta-decay half-lives of $N=50$ waiting point nuclei, starting from chiral NN and 3N interactions and consistent currents, without phenomenological adjustments. The available experimental half-lives are well described once 2B currents are included. Our exploratory study of FF contributions suggests that they are smaller below $Z=27$. This work shows that {\it ab initio} calculations can provide important input for astrophysics applications and guidance for beta-decay experiments at the neutron-rich extremes. Future work should include more detailed studies of the EFT interaction uncertainties and a consistent inclusion of FF contributions with 2B currents as well.

\begin{acknowledgments}
We thank G.~Mart\'inez-Pinedo, F.~Minato, T.~Neff, and S.R.~Stroberg for useful discussions.
This work was supported in part by the European Research Council (ERC) under the European Union’s Horizon 2020 research and innovation programme (Grant Agreement No.~101020842), the JST ERATO Grant No.~JPMJER2304, Japan, and by JSPS KAKENHI Grant No. 25K07294, No. 25K00995, and No. 25K07330. The authors gratefully acknowledge the Gauss Centre for Supercomputing e.V. for funding this project by providing computing time through the John von Neumann Institute for Computing (NIC) on the GCS Supercomputer JUWELS at J\"{u}lich Supercomputing Centre (JSC). Part of the numerical calculations were performed with the resources provided by Multidisciplinary Cooperative Research Program in Center for Computational Sciences, University of Tsukuba.
\end{acknowledgments}

\textit{Data availability}---The data that support the findings of this article are openly available~\cite{N50ZenodoData}.

\bibliography{ref}

\end{document}